# Evaluating Deep Surrogate Models for Knee Joint Contact Mechanics Under Input-Limited Conditions

Zhengye Pan[a,b], Jianwei Zuo[a,b], Jiajia Luo[*a,b]

[a] Biomedical Engineering Department, Institute of Advanced Clinical Medicine, Peking University, Beijing, China

[b] Institute of Medical Technology, Peking University Health Science Center, Peking University, Beijing, China

[*] Corresponding author, jiajia.luo@pku.edu.cn

**Abstract**:

**Background and Objective**: Accurate surrogate modeling of knee joint contact mechanics is important for reconstructing stress distributions and identifying risk-relevant regions, yet the relative suitability of different modeling paradigms under practically relevant input-limited conditions remains unclear.

**Methods**: Nine male soccer players performed 90° change-of-direction trials. Finite element simulations driven by subject-specific joint posture and reaction forces were converted into graph-structured samples. Five surrogate architectures representing local diffusion, history-context enhancement, hierarchical multi-scale modeling, explicit global interaction, and local-global hybridization were compared using three-fold cross-subject validation under full, pose-corrupted, load-corrupted, and minimal-input conditions. Performance was evaluated using full-field error, high-stress error, high-risk region overlap, and hotspot localization metrics.

**Results**: The hybrid model achieved the best overall performance under full inputs and remained the most robust under pose- and load-corrupted conditions. Under minimal inputs, no single model dominated all metrics: the history-context model yielded lower overall and high-stress errors, the hybrid model better preserved high-risk region reconstruction, and the hierarchical model showed an advantage in hotspot localization.

**Conclusion**: Evaluation of surrogate models for knee joint contact mechanics should shift from accuracy comparisons under ideal inputs to a comprehensive assessment of the preservation of risk-relevant information under realistic input constraints. Although the local-global hybrid model showed the best overall robustness, the optimal model under minimal-input conditions remained task-dependent.

# 1 Introduction

The knee joint undergoes complex multi-tissue contact and load transfer during movement. The resulting internal stress distribution, particularly the emergence of high-stress regions and hotspots, is closely linked to meniscal injury, articular cartilage degeneration, and the progression of osteoarthritis(Segal et al. 2012; Englund 2009). Accurate characterization of knee joint contact mechanics is therefore essential for elucidating injury mechanisms and identifying high-risk mechanical environments(Andriacchi et al. 2004). Finite element analysis (FEA), with its ability to represent joint geometry, material properties, and contact behavior with high fidelity, has become a key tool for investigating knee joint contact mechanics(Cooper et al. 2019; Zhang et al. 2025). However, its practical use is constrained by complex model construction, high computational cost, and heavy reliance on specialized expertise, thereby limiting its suitability for large-cohort studies, rapid screening, and near-real-time applications(Paz et al. 2021).

Against this background, deep surrogate models built on finite element meshes have emerged as a promising strategy for rapidly mapping upstream mechanical inputs, such as joint posture and loading, to internal stress field distributions(Eskinazi and Fregly 2015; Liang et al. 2018). In recent years, surrogate modeling of knee joint contact mechanics has advanced substantially, and the introduction of various network architectures and modeling mechanisms has markedly improved predictive accuracy under ideal input conditions(Ma et al. 2025). However, relatively little attention has been paid to how different surrogate-modeling paradigms perform under practically relevant input constraints(Zhao et al. 2024; Tan et al. 2023). In practice, ideal inputs are rarely available. On the one hand, posture measurement errors, kinematic reconstruction biases, and uncertainty in load estimation commonly cause posture- and load-related inputs to deviate from their true values, leading to unavoidable degradation in input quality(Camomilla et al. 2017). On the other hand, in clinical settings, during sideline monitoring, or in low-cost sensing environments, complete dynamic information is often unavailable, and models must operate with simplified or partially missing inputs(Bissenov et al. 2025). Accuracy under complete, high-quality inputs is therefore no longer sufficient as the sole basis for model evaluation. The more practically relevant question is which modeling paradigms can still reconstruct the overall stress distribution while preserving key risk-relevant information, including the high-stress tail, high-risk regions, and hotspot localization, under input-limited conditions(Farrokhi et al. 2011; Besier et al. 2008).

To address this question, the present study systematically evaluates surrogate models for knee joint contact mechanics under input-limited conditions that are relevant to practical use. Change-of-direction (CoD), a task characterized by pronounced high-stress exposure and richer risk-relevant spatial patterns, was selected as the representative scenario(Dos'Santos et al. 2018). Five representative paradigms were systematically compared: local topology diffusion, history-context enhancement, hierarchical multi-scale modeling, explicit global interaction, and local-global hybrid modeling. Particular emphasis was placed on their suitability under two typical forms of input limitation: input-quality degradation and input simplification. The former refers to numerical deviations in inputs such as joint posture or loading, whereas the latter represents situations in which complete upstream information is insufficient or partially unavailable. Within a unified training framework, scenario-specific evaluations were conducted to compare these paradigms with respect to full-field reconstruction, preservation of the high-stress tail, agreement in high-risk regions, and hotspot localization. The aim was to move the evaluation of surrogate models for knee joint contact mechanics beyond accuracy comparisons under ideal inputs toward a benchmark-based assessment of how well risk-relevant information is preserved under practical input constraints.

## 2 Methods

**2.1 Dataset construction**

2.1.1 Participant recruitment and data acquisition

A total of nine adult male soccer players were recruited for this study (174.4 ± 4.3 cm; 72.43 ± 7.1 kg; 22.1 ± 1.7 years). Eligible participants had no history of knee pain or knee surgery, no evident lower-limb motor dysfunction, and a dominant right leg. All participants were instructed to refrain from competition for 48 h before testing. Before participation, each subject was informed of the study purpose and procedures and provided written informed consent. The study was approved by the Ethics Approval Committee of Beijing Normal University (Approval No.TY202212088).

Kinematic data were collected using a Vicon three-dimensional motion capture system (200 Hz; V5, Oxford, UK), and ground reaction force (GRF) data were synchronously recorded using a Kistler force platform (1000 Hz; 9286AA, Winterthur, Switzerland). The reflective marker placement protocol followed Pan et al. (2024). Before formal testing, each participant completed a 5-min warm-up consisting of jogging (2.5 m/s) and self-directed stretching. During testing, participants performed five 90° CoD trials from the same standardized preparatory posture with an approach speed of $5 \pm 0.5$ m/s. The three trials with approach speeds closest to the target value and with full foot contact on the force platform were retained for analysis. An infrared timing gate was positioned 5 m from the center of the force platform to monitor approach speed, with the gate height aligned to the participant's greater trochanter. Initial contact was defined as the instant when vertical GRF exceeded 20 N, and toe-off as the instant when vertical GRF fell below 20 N; the interval between these two events was defined as the stance phase.

2.1.2 Data processing

The recorded kinematic marker trajectories and GRF data were filtered using a fourth-order zero-lag Butterworth low-pass filter with a cutoff frequency of 10 Hz. In OpenSim, the knee joint of the Rajagopal full-body musculoskeletal model was modified to include three explicit rotational degrees of freedom: flexion-extension, internal-external rotation, and abduction-adduction(Rajagopal et al. 2016). Inverse kinematics, inverse dynamics, static optimization, and joint reaction analysis were then performed sequentially to obtain the knee joint posture and joint reaction forces for finite element analysis. These variables were expressed in a unified tibial anatomical coordinate system and used as boundary-condition inputs for the subsequent finite element analysis.

2.1.3 Finite element simulation and dataset generation

A single generic knee joint finite element model from OpenKnee(s) was adopted as the unified reference model(Chokhandre et al. 2023). All participants shared the same anatomical geometry and mesh topology, and inter-subject differences arose solely from the subject-specific knee joint posture trajectories and joint reaction force trajectories obtained from musculoskeletal analysis. Finite element simulations were conducted in FEBio within a quasi-static framework using a time step of 0.01.

For boundary-condition assignment, the knee joint posture output from OpenSim was defined as the three rotational components of the femur relative to the tibia, namely flexion-extension, internal-external rotation, and abduction-adduction, and these were imposed directly as kinematic constraints in the finite element model. Translational degrees of freedom were not prescribed in advance as explicit kinematic inputs; instead, they were determined jointly by the applied joint reaction forces, tissue contact interactions, and mechanical equilibrium during solution. Meanwhile, the joint reaction forces computed in OpenSim were applied to the distal femoral reference node as a full three-dimensional load vector, including compressive, anterior-posterior, and medial-lateral components. This reference node was rigidly coupled to the femur and served as the control point through which the net joint load was transmitted, thereby reconstructing boundary-loading conditions corresponding to the measured movement.

Articular cartilage was modeled using an isotropic Mooney-Rivlin hyperelastic constitutive law, whereas the menisci and ligaments were modeled using a transversely isotropic Mooney-Rivlin constitutive law to represent their fiber-reinforced characteristics; initial ligament tension was introduced through prestrain. The material parameters for all tissues are listed in Table 1. Frictionless contact was defined between all articulating surfaces.

**Table 1.** Material coefficients used in the finite element model.

| Tissue | C1 (MPa) | C2 (MPa) | K (MPa) | C3 (MPa) | C4 | C5 (MPa) | m |
|---|---|---|---|---|---|---|---|
| Articular cartilage | 2.54 | 0 | 100 | - | - | - | - |
| Meniscus | 4.61 | 0 | 92.16 | 0.1197 | 150.0 | 400.0 | 1.019 |
| ACL | 1.95 | 0 | 146.41 | 0.0139 | 116.22 | 535.039 | 1.046 |
| PCL | 3.25 | 0 | 243.90 | 0.1196 | 87.178 | 431.063 | 1.035 |
| MCL | 1.44 | 0 | 793.65 | 0.57 | 48.0 | 467.1 | 1.063 |
| LCL | 1.44 | 0 | 793.65 | 0.57 | 48.0 | 467.1 | 1.063 |
| Patellar ligament | 2.75 | 0 | 206.61 | 0.065 | 115.89 | 777.56 | 1.042 |
| Quadriceps tendon | 2.75 | 0 | 206.61 | 0.065 | 115.89 | 777.56 | 1.042 |

Each time step of the finite element simulation was converted into a graph-structured sample to construct the dataset for surrogate model training. For each finite element time step, the graph sample was defined as $G = (V, E)$, where each graph node $v_i \in V$ corresponded to the centroid of a finite element rather than to a finite element mesh node, and each edge $e_{ij} \in E$ represented an adjacency relation in the mesh topology. Node features comprised the initial geometric position of each element together with the global conditioning variables at the corresponding time step. The global conditioning variables, including joint posture and loading, were broadcast to all graph nodes. Edge features were defined as the relative position vectors between adjacent graph nodes. For the target von Mises stress, a combined logarithmic transformation and Z-score normalization strategy was adopted to balance the numerical distribution and improve graph neural network training.

## 2.2 Deep surrogate models

### 2.2.1 Network architectures

To compare the behavior of different modeling paradigms under input-limited conditions, five representative network architectures were designed to span the spectrum from local diffusion to global interaction and from single-mechanism to composite-mechanism modeling, namely local topology diffusion, history-context enhancement, hierarchical multi-scale modeling, explicit global interaction, and a local-global hybrid model (Figure 1). The models were matched as closely as possible in terms of input-output definitions, training pipelines, backbone hidden dimensions, and basic propagation depth, with the main differences concentrated in their core propagation and information-interaction mechanisms. All models followed a common encoder-core propagation module-decoder framework: the encoder mapped input features into a latent space, the core propagation module performed information updating and graph interactions, and the decoder generated stress predictions. For brevity, the five architectures are hereafter denoted as MGN, CT, Hi, GI, and Hy, respectively.

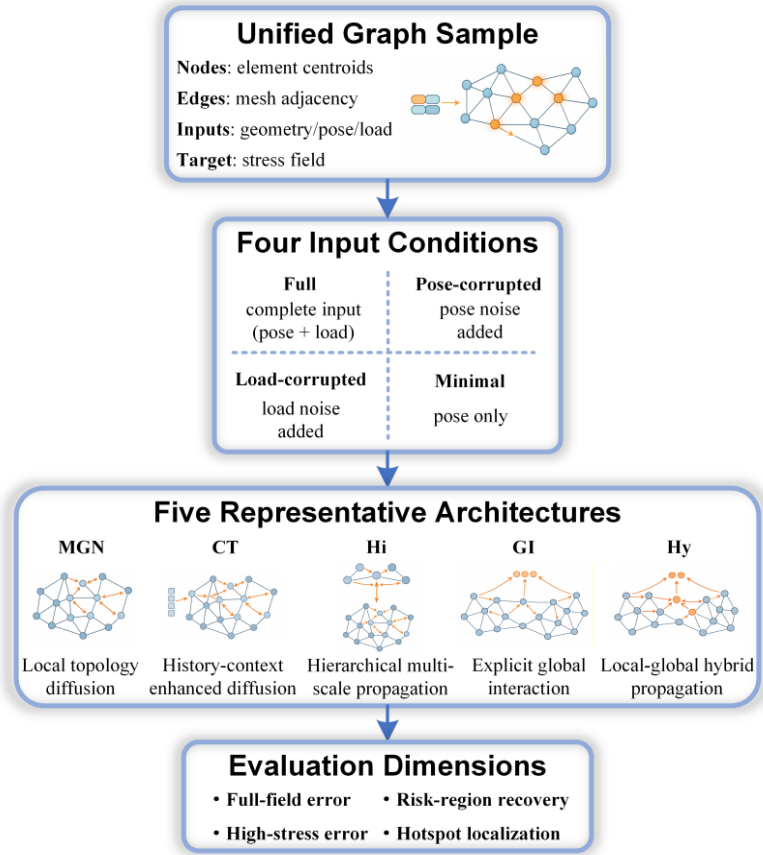

**Figure 1.** Overall workflow for comparing five surrogate architectures under four input conditions. Finite element results were converted into unified graph samples, and five representative surrogate architectures were compared under four input conditions (Full, Pose-corrupted, Load-corrupted, and Minimal) using full-field error, high-stress error, high-risk region reconstruction, and hotspot localization as evaluation dimensions.

The local topology diffusion model corresponds to the standard MeshGraphNet (MGN) (Pfaff et al. 2021). Built on the adjacency structure of the finite element mesh, it performs layer-wise message passing along edges between neighboring graph nodes, thereby enabling local feature diffusion and state updating. This model directly exploits mesh connectivity to characterize stress propagation between adjacent tissue elements and therefore serves as the most fundamental comparison paradigm in the present study. Because long-range dependencies are captured primarily through the gradual accumulation of information across multiple propagation layers, this provides a baseline for evaluating the modeling capacity of local diffusion mechanisms.

The history-context enhancement model (CT) extends the local topology diffusion framework by incorporating short-term historical sequences to supplement recent boundary-condition changes that cannot be fully captured from a single-frame input. Specifically, the model encodes joint posture and loading from several preceding time steps and injects the resulting contextual representation into the feature-propagation process at the current time step, thereby introducing recent temporal cues beyond the instantaneous input alone. The historical information is not used to explicitly model inertial effects, viscoelasticity, or other dynamic mechanisms; rather, it serves to represent the contextual information carried by recent changes in posture and loading.

The hierarchical multi-scale model (Hi) constructs a coarse-grained hierarchical representation on top of the original graph structure to enable information aggregation and interaction across scales. Fine-grained graph nodes are first pooled to form region-level representations, propagation and updating are then performed in the coarse-scale space, and the resulting high-level information is finally passed back to the original scale. Compared with single-scale local diffusion, this design can integrate structural context over a broader range with fewer propagation steps, thereby strengthening the representation of long-range topological dependencies and global mechanical patterns.

Unlike the three architectures described above, the explicit global interaction model (GI) does not rely solely on layer-by-layer diffusion along the mesh topology to build a global receptive field. Instead, it establishes long-range information links directly through an explicit global communication mechanism. This design allows graph nodes at different spatial locations to exchange information across regions through a global channel, providing a more direct means of modeling long-range interactions and overall coupling relationships.

On this basis, the present study further combined local topology diffusion and explicit global interaction to construct a local-global hybrid model (Hy). The model retains local message passing along mesh adjacency relations to preserve local structural constraints, while introducing a global communication module to model long-range dependencies. These two information streams are fused in the latent space and jointly participate in state updating, thereby balancing local consistency and global integration.

2.2.2 Training and inference

All models were trained and evaluated under a unified data-splitting protocol and training strategy. The dataset was partitioned using a three-fold cross-subject scheme at the participant level: the nine participants were divided into three non-overlapping test folds, and in each fold, three participants were assigned to the test set while the remaining six formed the training pool. Within the training pool, a validation set was defined separately for each participant in temporal order; specifically, after sorting each participant's samples by time index, the final 10% were used for validation, and the remaining samples for training. All models shared the same training, validation, and test splits to avoid additional bias introduced by differences in data partitioning.

To prevent hyperparameter tuning from introducing test-set information leakage during cross-validation, model hyperparameters were determined through lightweight preliminary experiments before the formal three-fold cross-validation, and the same backbone configuration and training strategy were fixed for all folds and all models. During training, all models used the AdamW optimizer with an initial learning rate of $1 \times 10^{-3}$, a batch size of 1, and a maximum of 120 epochs. Learning-rate scheduling followed a ReduceLROnPlateau strategy: when the validation loss failed to improve for 10 consecutive epochs, the learning rate was multiplied by 0.5. Early stopping was also applied, such that training was terminated if the validation loss did not decrease for 30 consecutive epochs. To improve training stability, gradient norm clipping was performed during backpropagation, with the maximum norm set to 1.0. For all models, the parameters corresponding to the lowest validation loss were selected as the final model for testing.

Except for paradigm-specific modules, all models shared the same backbone hidden dimensionality ($D_{hidden} = 32$) and propagation depth ($K = 3$) to maintain consistency in backbone capacity and training budget. The history-context enhancement model used a short-term history window of length 8 to encode preceding conditioning information; when insufficient preceding time steps were available at the beginning of a sequence, the earliest available frame was replicated as padding to maintain a consistent history length. Because modules such as the history encoder, global interaction block, and hierarchical pooling block introduce additional structure, the models were not strictly matched in total parameter count or effective capacity. Accordingly, the comparison focused on the relative performance of different inductive biases under a unified backbone configuration and training budget.

The training objective was defined as the masked mean squared error (masked MSE) over the valid graph nodes, constraining the discrepancy between the predicted stress and the normalized target stress. During inference, the test set was first evaluated under the standard input condition (Full) to establish each model's baseline performance with complete inputs. Two additional types of representative input-limited conditions were then constructed on the same test set to assess robustness under more realistic deployment settings.

The first type was input-quality degradation, including two scenarios: posture corruption (Pose-corrupted) and load corruption (Load-corrupted). Specifically, zero-mean Gaussian noise was added to either the posture or load channels in the standardized feature space, with a standard deviation of 0.20 in both cases. This value was chosen to represent a moderate perturbation after standardization, large enough to induce measurable corruption while still preserving the basic input structure. The resulting perturbation was intended to mimic measurement errors or estimation biases that may arise in practical use settings, while avoiding unrealistically severe corruption. For the history-context enhancement model, the same perturbations were applied simultaneously to both the current input and the corresponding historical conditioning sequence to preserve temporal consistency.

The second type was input simplification (Minimal), corresponding to a minimal-input scenario in which posture information was retained. At the same time, the load channels were set to zero, thereby simulating operation when complete upstream dynamic information was unavailable. For the history-context enhancement model, the load channels in the historical conditioning sequence were likewise set to zero in a synchronized manner. To avoid additional variability introduced by random perturbations across evaluation modes, all test-time perturbations were generated independently for each sample under a fixed random seed and were kept identical across models.

2.2.3 Evaluation metrics

To systematically evaluate predictive performance under both complete-input and input-limited conditions, model outputs were assessed from four perspectives: full-field error, high-stress error, high-risk region reconstruction, and hotspot localization. First, root mean squared error (RMSE) and mean absolute error (MAE) were used to quantify the reconstruction accuracy of the overall stress field and the overall deviation between predictions and ground truth. In addition, the Pearson correlation coefficient r was used to assess the linear agreement between the predicted and true stress fields in terms of their spatial distributions.

To evaluate each model's ability to preserve the magnitude characteristics of high-stress regions, the relative error of maximum stress ($RE_{max}$) and the relative error at the 95th percentile

($RE_{P95}$) were calculated. $RE_{max}$ reflects the ability to reconstruct the peak stress, whereas $RE_{P95}$ characterizes prediction bias in the high-stress tail of the distribution. Because knee-joint injury risk is often closely associated with localized high-stress exposure, these two metrics provide a complementary assessment of performance with respect to risk-relevant stress magnitudes(Besier et al. 2015).

High-risk regions were defined as the set of graph nodes with stress values above the 95th percentile of the stress field, and the Dice coefficient was used to evaluate spatial overlap between the predicted and true high-risk regions. To assess hotspot localization, the hotspot localization error, $d_{hot}$, was defined as the Euclidean distance between the centroids of the predicted and true high-risk regions. This metric further characterizes the bias in the spatial localization of the high-risk region center and complements the Dice coefficient, which describes the degree of regional overlap.

## 2.3 Statistical analysis

All evaluation metrics were first summarized at the participant level; specifically, for each participant, all test samples under a given model and input condition were averaged to obtain participant-level metric values. Statistical analyses were then performed on these participant-level results. All results are presented as mean ± standard deviation. For each evaluation metric under each input condition, a Friedman test was first conducted to assess whether an overall difference existed among the five model types. When the Friedman test was statistically significant, post hoc pairwise comparisons were performed using two-sided Wilcoxon signed-rank tests, and the resulting p-values were adjusted for multiple comparisons using the Holm-Bonferroni method. The significance level was set at $\alpha = 0.05$.

# 3 Results

## 3.1 Performance Comparison Under the Full Condition

The comparisons of the five models in terms of full-field error, high-stress error, high-risk region reconstruction, and hotspot localization under the Full condition are shown in Figure 2. Overall, Hy achieved the best performance in RMSE, $RE_{max}$, and Dice, significantly outperforming the other models ($p < 0.05$). GI showed the weakest performance on these three metrics, with RMSE and $RE_{max}$ significantly higher than those of Hy ($p < 0.05$) and Dice significantly lower than that of

Hy ($p < 0.05$). MGN, Hi, and CT generally occupied the intermediate range. Among them, CT was numerically slightly better than MGN and Hi in RMSE and $RE_{max}$, whereas the three models showed only minor differences in Dice. For the hotspot localization error, $d_{hot}$, Hy remained relatively low, whereas GI tended to be higher; however, between-model separation was weaker for $d_{hot}$ than for RMSE, $RE_{max}$, and Dice.

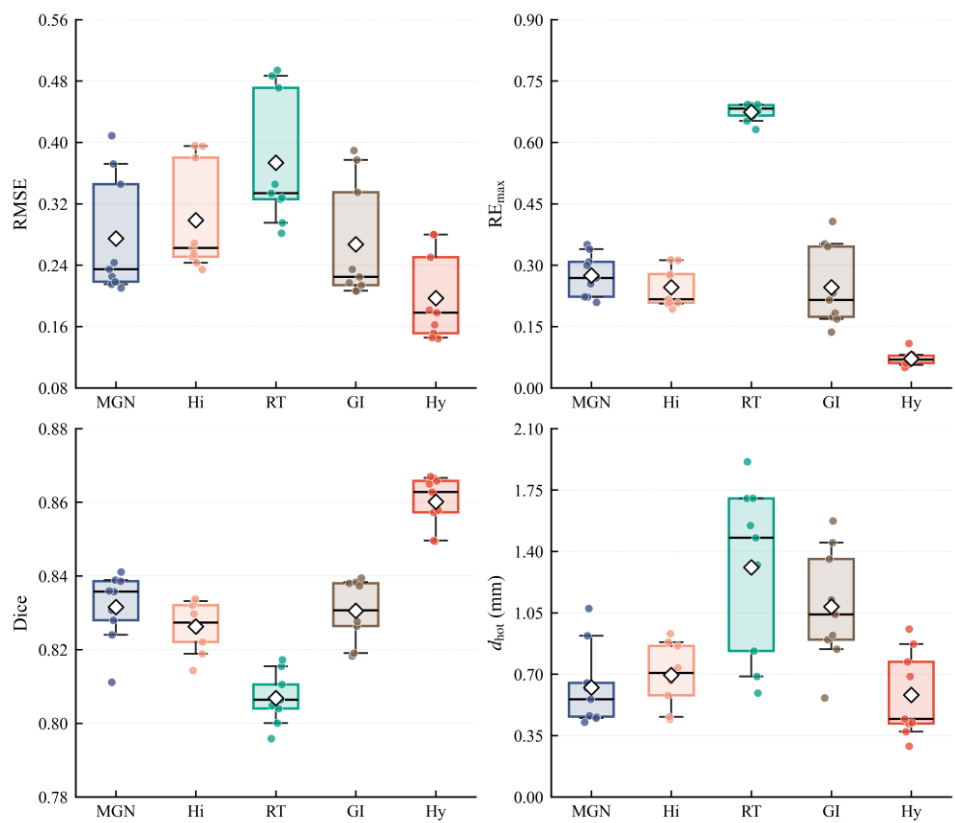

**Figure 2.** Comparison of the five models in RMSE, $RE_{max}$, Dice, and $d_{hot}$ under the Full condition. Colored dots represent individual subject-level values with slight horizontal jitter for visualization; hollow diamonds indicate the mean; boxes indicate the interquartile range; center lines indicate the median; and whiskers indicate the 5$^{th}$ and 95$^{th}$ percentiles.

Table 2 further presents the results of the five models under the Full condition in terms of spatial agreement, absolute error, and high-stress tail error. Overall, Hy performed strongly in both $r$ and MAE, achieving high spatial correlation and low mean absolute error, respectively ($p < 0.05$). For $RE_{P95}$, both Hy and MGN remained relatively low, whereas GI and CT were higher. In addition to Hy, MGN and CT also exhibited comparatively high $r$ values, whereas GI was characterized by lower r and higher MAE.

**Table 2.** Comparison of the five models in *r*, MAE, and RE$_{P95}$ under the Full condition.

| Model | *r* | MAE | RE$_{P95}$ |
|---|---|---|---|
| MGN | 0.953±0.007 | 0.096±0.027 | 0.012±0.003 |
| Hi | 0.944±0.004 | 0.097±0.023 | 0.016±0.006 |
| GI | 0.914±0.005 | 0.105±0.026 | 0.017±0.008 |
| CT | 0.956±0.007 | 0.098±0.028 | 0.019±0.008 |
| Hy | **0.975±0.005** | **0.076±0.020** | 0.011±0.003 |

**Note:** *r*, Pearson correlation coefficient; MAE, mean absolute error; RE$_{P95}$, relative error at the 95th percentile stress. MGN, MeshGraphNet; Hi, hierarchical multi-scale propagation; GI, explicit global interaction; CT, history-context enhanced diffusion; Hy, local-global hybrid propagation. Bold indicates that the corresponding metric under the given evaluation setting was significantly better than that of all other models ($p < 0.05$).

## 3.2 Performance Comparison Under Input-Limited Conditions

The performance trends of the five models under different input-limited conditions in terms of full-field error, high-stress error, high-risk region reconstruction, and hotspot localization are shown in Figure 3. Overall, all models deteriorated to varying degrees when moving from the Full condition to input-limited conditions. The deterioration was most pronounced under the Minimal condition, as reflected by overall increases in RMSE and $d_{hot}$ and a marked decrease in Dice. In contrast, the changes under the Pose-corrupted and Load-corrupted conditions were relatively modest, and for most models, the degradation under Load-corrupted was less severe than that under Pose-corrupted.

Across the Full, Pose-corrupted, and Load-corrupted conditions, Hy consistently maintained lower RMSE and RE$_{max}$ values and higher Dice values, and significantly outperformed the other models in all three settings ($p < 0.05$). By contrast, GI performed worst under these conditions, particularly with persistently high RE$_{max}$ values. MGN, Hi, and CT generally fell in the intermediate range. Among them, CT showed a more pronounced increase in RMSE under the Pose-corrupted condition, whereas its overall performance partially recovered under the Load-corrupted condition. Under the Minimal condition, however, the models no longer exhibited a consistent overall ranking; instead, they showed differentiated strengths across metrics: CT tended to have lower RMSE and RE$_{max}$, Hy showed higher Dice, and Hi yielded lower $d_{hot}$.

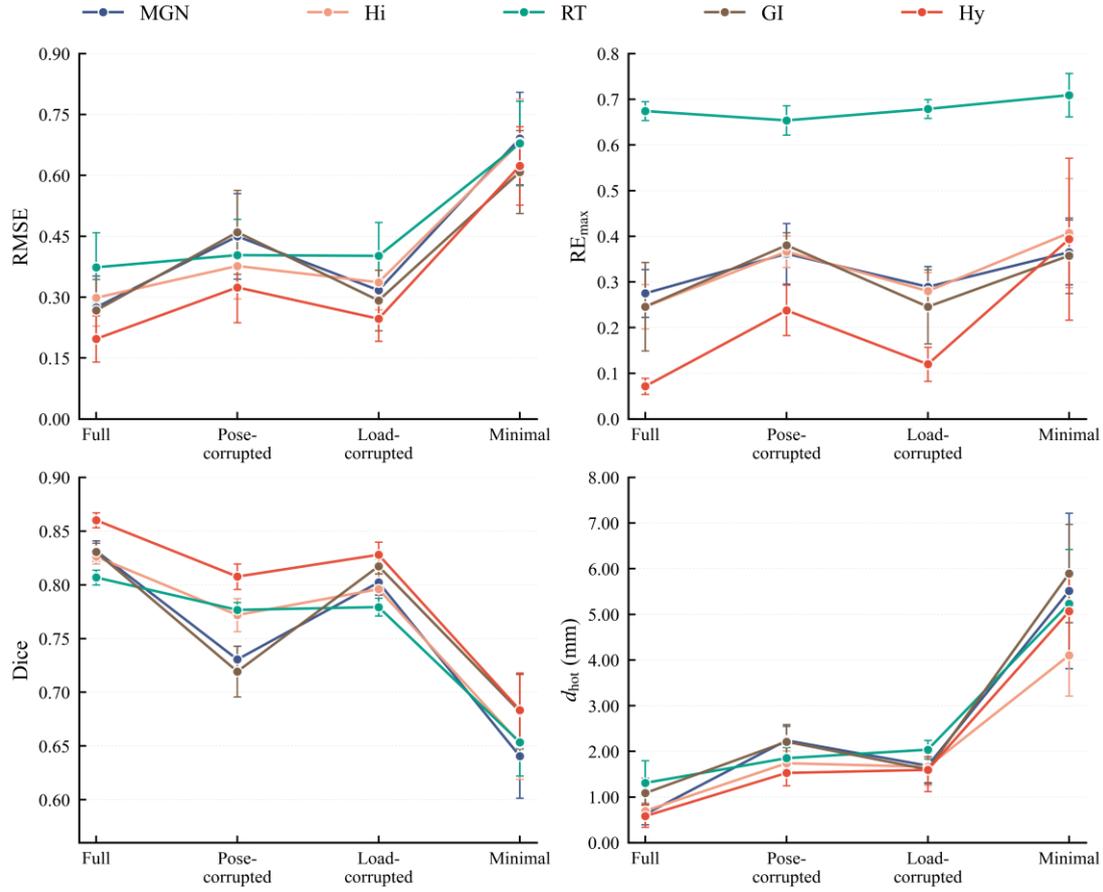

**Figure 3.** Comparison of the five models in RMSE, RE$_{max}$, Dice, and $d_{hot}$ under different input-limited conditions. Points indicate the mean across subject-level values, and error bars denote the standard deviation. RMSE, root mean squared error; RE$_{max}$, relative error of maximum stress; $d_{hot}$, hotspot localization error. MGN, MeshGraphNet; Hi, hierarchical multi-scale propagation; GI, explicit global interaction; CT, history-context enhanced diffusion; Hy, local-global hybrid propagation.

Table 3 presents the results of the five models for $r$ and RE$_{P95}$ under different input-limited conditions. Overall, spatial agreement was higher under the Load-corrupted condition than under the Pose-corrupted and Minimal conditions, whereas under the Minimal condition, $r$ declined more noticeably, and RE$_{P95}$ increased overall. Under the Pose-corrupted condition, Hy showed relatively high $r$, whereas GI exhibited comparatively low RE$_{P95}$. Under both the Load-corrupted and Minimal conditions, Hy maintained relatively high $r$ values, while CT showed comparatively low RE$_{P95}$ values.

**Table 3.** Comparison of the five models in $r$ and $RE_{P95}$ under different input-limited conditions.

|  | Pose-corrupted | | Load-corrupted | | Minimal | |
|---|---|---|---|---|---|---|
|  | $r$ | $RE_{P95}$ | $r$ | $RE_{P95}$ | $r$ | $RE_{P95}$ |
| MGN | 0.870±0.010 | 0.041±0.007 | 0.942±0.007 | 0.057±0.009 | 0.811±0.032 | 0.320±0.061 |
| Hi | 0.910±0.007 | 0.037±0.005 | 0.933±0.005 | 0.057±0.009 | 0.822±0.021 | 0.318±0.065 |
| GI | 0.899±0.003 | 0.036±0.004 | 0.904±0.004 | 0.057±0.008 | 0.816±0.023 | 0.302±0.058 |
| CT | 0.862±0.017 | 0.044±0.007 | 0.950±0.007 | 0.052±0.010 | 0.863±0.024 | 0.285±0.066 |
| Hy | **0.934±0.015** | 0.037±0.004 | **0.965±0.005** | 0.056±0.010 | 0.870±0.026 | 0.311±0.057 |

**Note:** $r$, Pearson correlation coefficient; $RE_{P95}$, relative error at the 95th percentile stress. MGN, MeshGraphNet; Hi, hierarchical multi-scale propagation; GI, explicit global interaction; CT, history-context enhanced diffusion; Hy, local-global hybrid propagation. Bold indicates that the corresponding metric under the given evaluation setting was significantly better than that of all other models ($p < 0.05$).

### 3.3 Temporal Performance Degradation During the Stance Phase

Figure 4 illustrates the increase in RMSE relative to the Full condition across the stance phase for the five models under the three input-limited conditions. Overall, all three input-limited conditions increased RMSE relative to the Full condition, but their temporal patterns differed markedly. Under the Pose-corrupted condition, all models showed a sustained increase in RMSE throughout the stance phase, with MGN and CT exhibiting larger increases overall and GI showing a smaller increase. As the stance phase progressed, the increase became more pronounced for most models during the later stage. Under the Load-corrupted condition, the increase in RMSE was generally small across models, with only mild variation over the stance phase and limited differences between models. By contrast, under the Minimal condition, the increase in RMSE was much more pronounced, rising rapidly during late stance, with a clear exacerbation emerging after approximately 70% of the stance phase. In this setting, MGN and Hy showed larger late-phase increases, whereas Hi exhibited a smaller increase.

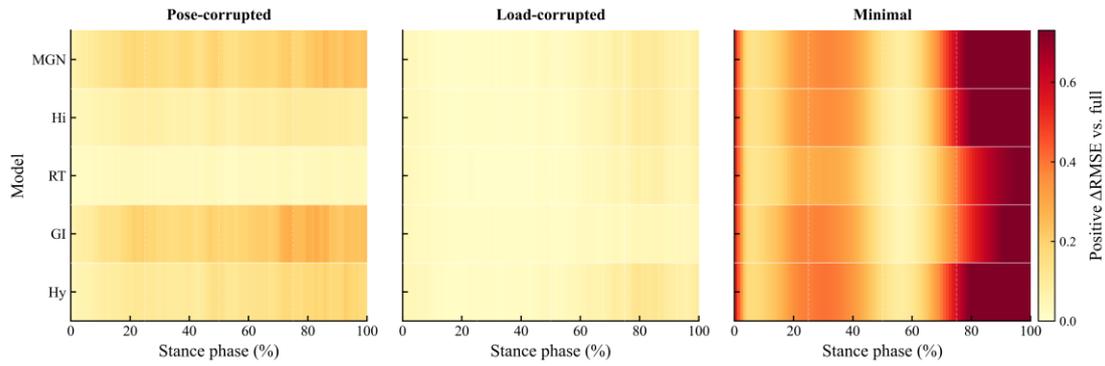

**Figure 4.** Heatmaps of phase-wise RMSE increase relative to the Full condition for the five models under the Pose-corrupted, Load-corrupted, and Minimal conditions. The horizontal axis represents the stance phase, and color indicates the positive ΔRMSE with respect to the Full condition. MGN, MeshGraphNet; Hi, hierarchical multi-scale propagation; GI, explicit global interaction; CT, history-context enhanced diffusion; Hy, local-global hybrid propagation.

Figure 5 shows the decrease in Dice relative to the Full condition across the stance phase for each model under the three input-limited conditions. Similarly, all three input-limited conditions reduced Dice relative to the Full condition. Under the Pose-corrupted condition, all models exhibited a sustained decrease in Dice throughout the stance phase, with CT and MGN showing larger declines overall. As the stance phase progressed, most models still showed a pronounced reduction during the later stage. Under the Load-corrupted condition, the decrease in Dice was generally small across models and was concentrated mainly in the late stance. In contrast, under the Minimal condition, the decrease in Dice was more substantial and persisted throughout the stance phase, with particularly marked reductions during the early and late stages. Under this condition, MGN, Hi, and Hy showed larger overall declines, whereas GI showed a smaller decline.

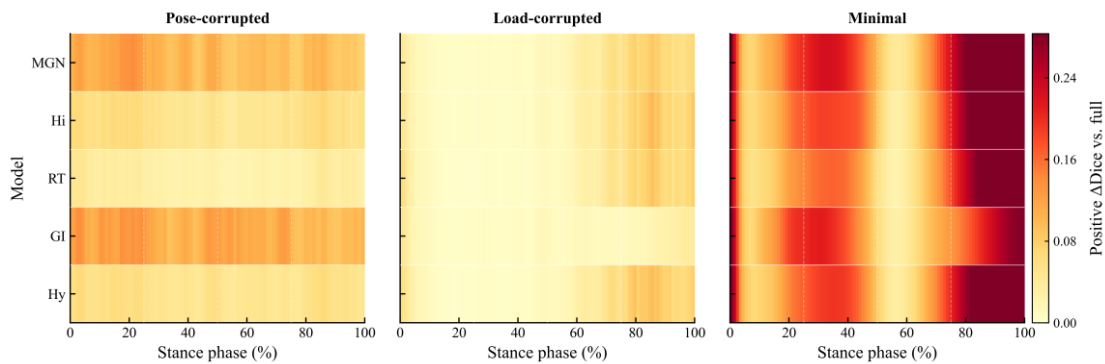

**Figure 5.** Heatmaps of phase-wise Dice decrease relative to the Full condition for the five models under the Pose-corrupted, Load-corrupted, and Minimal conditions. The horizontal axis represents

the stance phase, and color indicates the magnitude of Dice reduction with respect to the Full condition. MGN, MeshGraphNet; Hi, hierarchical multi-scale propagation; GI, explicit global interaction; CT, history-context enhanced diffusion; Hy, local-global hybrid propagation.

## 4 Discussion

This study systematically compared five representative paradigms for surrogate modeling of knee joint contact mechanics under the practical demands of real-world application, evaluating their performance under the Full condition and three input-limited conditions. The results showed that under the Full condition, Hy performed best overall across most metrics. Under the Pose-corrupted and Load-corrupted conditions, Hy still maintained stable overall performance, and the degradation from posture perturbation was generally greater than that from load perturbation. Under the Minimal condition, however, the models exhibited distinct strengths across performance dimensions: CT yielded lower overall error and higher-stress-magnitude error, Hy performed better in high-risk region reconstruction, and Hi showed lower hotspot localization error. These findings suggest that the evaluation of surrogate models for knee joint contact mechanics should not be confined to accuracy comparisons under complete-input conditions, but should place greater emphasis on preserving risk-relevant information under input-limited conditions(Hicks et al. 2015).

Under the Full condition, Hy performed well across multiple metrics, suggesting that the combination of local topology diffusion and explicit global interaction may offer complementary advantages(Patrignani and Pinho 2026). The knee joint contact stress field is governed not only by local load transfer between adjacent tissue elements, but also by the combined effects of global geometric relationships, redistribution of contact states, and cross-region coupling. When relying solely on local diffusion, long-range dependencies must be accumulated gradually through multiple propagation layers; when relying solely on global communication, the local structural constraints provided by mesh topology may be weakened. The superior performance of Hy across multiple metrics indicates that, for mechanical problems such as those examined here, which involve both local contact behavior and global load coupling, a single inductive bias is often insufficient to support both full-field reconstruction and preservation of risk-relevant regions. In contrast, the synergy between local and global mechanisms is more advantageous.

Compared with the Full condition, the input-quality degradation conditions more closely resemble error sources encountered in real-world deployment and therefore provide a more meaningful reflection of practical model utility. In the present study, performance degradation under Pose-corrupted conditions was generally greater than that under Load-corrupted conditions, suggesting that the models were more sensitive to errors in posture-related inputs. One possible explanation is that posture affects not only stress magnitude but also directly determines the spatial organization of contact regions, local load-transfer pathways, and the overall pattern of high-stress distribution. In contrast, load perturbations may act primarily at the level of stress magnitude, with a more limited effect on spatial patterns(Liao et al. 2018). However, this interpretation is based on results obtained under a unified geometry and quasi-static finite element framework. It should not be generalized as a universal conclusion regarding the relative biomechanical importance of posture and loading.

The performance divergence observed under the Minimal condition is a notable finding of this study. When complete loading information was unavailable, the models no longer exhibited a consistent overall ranking; instead, they showed differentiated strengths across performance dimensions. This result suggests that, in minimal-input scenarios, model suitability may depend on the specific type of target information prioritized. Based on the present results, history-context enhancement was more consistently associated with the preservation of overall error and high-stress-magnitude reconstruction; the local-global hybrid mechanism was more consistently associated with the preservation of the spatial pattern of high-risk regions; and hierarchical multi-scale representation showed potential for hotspot-center localization. This finding further indicates that model selection for real-world applications should no longer rely on a single accuracy ranking derived from ideal conditions, but should instead be matched to the type of risk-relevant information that must be preserved.

The temporal analysis of the stance phase further showed that the degradation due to input limitations was not uniformly distributed throughout the movement. Degradation under the Load-corrupted condition was generally mild, whereas that under the Pose-corrupted and Minimal conditions was more pronounced, with the Minimal condition showing marked exacerbation during late stance. This finding suggests that when upstream input information is substantially insufficient, predictive stability is more likely to be compromised during the later stage of the movement. For high-risk tasks such as change-of-direction maneuvers, this has practical significance because what matters in real deployment is not merely the average error over the entire movement but also whether the model can stably retain its ability to identify abnormal mechanical environments during the most critical high-risk periods.

This study has several limitations. First, the sample size was small, and both the participant cohort and movement task were relatively homogeneous; therefore, the current findings require further validation in larger samples and across a broader range of movement scenarios. Second, the study adopted a unified training and scenario-specific testing design to better reflect the deployment-oriented logic of evaluating already trained models in real-world settings; however, this does not represent the upper performance bound each model might achieve if retrained specifically for input-limited conditions. Finally, the present work still focuses primarily on finite element stress field reconstruction as the core endpoint and has not yet been extended to clinical outcomes or injury-event prediction. Future studies should expand both sample size and task diversity, conduct more comprehensive evaluations under multi-level input-limited conditions, and link risk-relevant mechanical representations with clinically observable endpoints, thereby advancing surrogate models from numerical replacement tools toward genuinely usable tools for risk assessment.

## 5 Conclusion

The evaluation of surrogate models for knee joint contact mechanics should shift from accuracy comparisons under ideal input conditions toward a more comprehensive assessment of their ability to preserve risk-relevant information under input-limited conditions. For the mechanical problems examined in this study, in which local contact behavior and global load coupling coexist, models that integrate local and global mechanisms were more suitable overall. When input information was substantially constrained, however, the relative performance of different models depended on the specific type of risk-relevant information prioritized by the intended application scenario.

## Declaration of Interest Statement

The authors have no conflicts of interest to declare.

## Funding Statement

This study was supported by the National Key R&D Program of China (grant no. 2023YFC2411201); Beijing Natural Science Foundation (grant no. L259081); NSFC General Program (grant no. 31870942).